# Forward Raman Scattering and Self-Modulation instabilities of lasers in magnetized tapered plasma channels

S. S. Ghaffari-Oskooei[1], A. A. Molavi Choobini[2*]

[1]Department of Atomic and Molecular Physics, Faculty of Physics, Alzahra University, Tehran, Iran,

[2]Dept. of Physics, University of Tehran, Tehran 14399-55961, Iran.

**Abstract:**

The propagation of laser pulses in tapered magnetized plasma channels is analyzed using the fluid theory of cold plasmas. This study focuses on laser propagation's key instabilities: forward Raman scattering and self-modulation instability. The influence of plasma density, laser intensity, dc magnetic field strength, and laser polarization on the growth rates of these instabilities is thoroughly examined. Analytical and numerical computations of the number of e-foldings for forward Raman scattering are performed to quantify its impact. The results reveal that increased plasma density and laser intensity significantly enhance the growth rates of these instabilities, leading to amplified Raman-scattered wave intensity and modulation of the laser envelope. Moreover, the strength of the dc magnetic field plays a pivotal role: it boosts instability growth rates for right-handed circularly polarized laser pulses while suppressing them for left-handed circularly polarized pulses. These findings highlight that reducing the growth rates of instabilities can facilitate the stable propagation of laser pulses in tapered plasma channels.

## I. Introduction

The propagation of high-intensity laser pulses in plasma channels is a topic of significant interest due to its critical role in various applications, such as inertial confinement fusion, second harmonic generation, X-ray lasers, and laser wakefield accelerators [1–5]. Achieving long-distance laser beam propagation is essential for these applications. Typically, the propagation distance of a laser beam is restricted to the Rayleigh length due to diffraction. However, by employing plasma channels, this distance can be extended to several tens of Rayleigh lengths. Various plasma channel designs have been proposed to guide laser pulses over long distances, including parabolic [6,7], leaky [8,9], single-mode [10], and tapered channels [11,12]. Tapered plasma channels, characterized by an inhomogeneous plasma density profile, are particularly suited for laser wakefield accelerators [11]. Beyond laser wakefield acceleration, tapered plasma channels also enable phase-locking of accelerating and focusing forces in plasma-based accelerators [13].

Despite their potential, the propagation characteristics of laser pulses in tapered plasma channels have been explored in only a limited number of studies. Previous work has examined the stability of intense pulses in plasma channels, considering factors such as group velocity dispersion, wakefields, and relativistic nonlinearities [12]. These studies demonstrated that tapered channels can effectively guide laser beams over extended distances. Hafizi et al. investigated the self-focusing of laser pulses in tapered plasma channels [14], concluding that appropriate selection of laser and plasma parameters enables self-focusing in such channels. However, laser pulses propagating through tapered plasma channels are prone to various instabilities, including Raman scattering, self-modulation, and Brillouin scattering [11]. These instabilities must be mitigated to ensure stable and efficient laser pulse propagation.

Forward Raman scattering (FRS) is a four-wave parametric instability involving the coupling of the incident laser beam with scattered waves (Stokes and anti-Stokes sidebands) and plasma waves [15–23]. This interaction is driven by the ponderomotive force, which amplifies both the plasma wave and the sidebands. The growth rate of FRS can be analytically derived using coupled wave equations for the laser, sidebands, and plasma waves [15,16]. Several studies have explored this instability under various conditions. For instance, Kamboj et al. analyzed the growth rate of FRS for vortex laser beams in plasma [18], while Xie et al. investigated the effects of Landau damping on FRS growth rates through Vlasov-Poisson simulations with different phase-space distributions [19]. Rozina et al. examined FRS in degenerate spin-polarized plasma using a quantum hydrodynamic model [20], and Wang et al. studied FRS in plasmas with sinusoidal density modulation [21]. Additionally, Tan et al. used two-dimensional particle-in-cell simulations to investigate stimulated Raman side scattering in plasmas [22]. Importantly, FRS generates plasma waves with high phase velocities, capable of accelerating electrons to relativistic speeds [17], making the study of this instability vital for plasma-based accelerator designs [23]. Self-modulation instability (SMI) is another phenomenon significantly impacting laser propagation in plasma. When a laser pulse propagates through plasma, it perturbs the plasma density, generating a plasma wave. The resulting density compression and rarefaction cause alternating focusing and defocusing of the laser, leading to intensity modulation. The ponderomotive force amplifies the plasma wave, driving the self-modulation instability [11,24]. Penano et al. conducted full-scale fluid simulations of SMI in tapered plasma channels, revealing interactions between FRS and SMI in plasma-based accelerators [11]. Yazdanpanah studied pulse evolutions driven by wakefield effects from SMI [25], while Chen et al. examined the growth rate of SMI in pair plasmas [26]. Esarey et al. analyzed the SMI of relativistic laser pulses in plasma by solving the beam envelope equation [28], and Krall et al. proposed using self-modulated laser pulses for alternative wakefield acceleration methods, demonstrating enhanced electron beam acceleration due to SMI [29].

This study investigates forward Raman scattering (FRS) and self-modulation instabilities (SMI) in magnetized tapered plasma channels analytically and numerically. The work introduces a tapered plasma density profile, defined by a plasma frequency that varies with position and channel radius. This innovative approach deviates from the traditional uniform plasma channels, enabling the analysis of spatially varying effects on FRS and SMI. Notably, the growth rate of FRS is derived with explicit dependence on spatial variables such as the plasma density gradient scale length and channel properties. The self-modulation growth rate derivation incorporates position-dependent parameters and introduces novel coupling terms, highlighting the influence of tapered channels on the evolution of instabilities. An external static magnetic field modifies the dispersion relation by introducing cyclotron frequency effects. The interplay between the cyclotron frequency and plasma parameters is analyzed, demonstrating how magnetization can be leveraged to control instability growth rates. Additionally, circularly polarized laser pulses add complexity to the nonlinear interaction. This polarization-dependent behavior, often overlooked in prior studies, is rigorously examined, providing new insights into the dynamics of laser-plasma interactions in magnetized, tapered environments. The paper is organized as follows: Section II derives the equations governing laser propagation in tapered plasma channels. Sections III and IV analyze FRS and SMI, respectively, in these channels. Numerical results are presented and discussed in Section V. Finally, conclusions are summarized in Section VI.

## II. General Mechanism

Consider a circularly-polarized laser pulse propagating in a tapered magnetized plasma channel immersed in an external static magnetic field $\vec{B}_0 = B_0 \hat{z}$. The vector potential of the laser pulse is expressed as:

$$\vec{A}(r,z,t) = \frac{1}{2} A(r,z,t) \exp(i\psi) (\hat{x} + i\sigma\hat{y}) + c.c. \tag{1}$$

where $\sigma = \pm 1$ indicates either right or left-handed circular polarization, respectively. $\psi$ is the phase of the vector potential given by:

$$\psi = \int_0^z k_0(z')dz' - \omega_0 t \tag{2}$$

The parameters of $\omega_0$ and $k_0$ are the frequency and wavenumber of the laser pulse, respectively. To simplify, the normalized vector potential $\vec{a} = \frac{-e}{mc^2}\vec{A}$ is introduced. The tapered plasma channel's density profile is defined as:

$$n(r,z) = n_0 \left(1 + \frac{z}{L}\right) \left(1 + \frac{r^2}{r_{ch}^2(z)}\right) \tag{3}$$

where the plasma density increases linearly with $z$ and the channel radius $r_{ch}$ is selected as follows:

$$r_{ch}(z) = \frac{\omega_p(z) r_0^2}{2c\sqrt{1 - \frac{Pe^2 \omega_p^2(z)}{2m^2 c^5 \omega_0^2}}} \tag{4}$$

here $\omega_p(z) = \sqrt{4\pi N(z) e^2/m}$, $P$ and $r_0$ are the laser power and spot size, respectively. The normalized vector potential satisfies Maxwell's wave equation:

$$\left(\nabla^2 - \frac{1}{c^2}\frac{\partial^2}{\partial t^2}\right)\vec{a} = \frac{4\pi e}{mc}\vec{J} \tag{5}$$

By substitution of the current density $\vec{J} = \frac{ecn(r,z)}{\gamma - \frac{\sigma\omega_c}{\omega_0}}\vec{a}$ and Lorentz factor $\gamma = \sqrt{1 + |a|^2/\left(1 - \frac{\sigma\omega_c}{\omega_0}\right)^2}$ in Eq. (5), one can derive the following equation in weakly-relativistic regime ($|a| \ll 1$):

$$(\nabla^2 - \frac{1}{c^2}\frac{\partial^2}{\partial t^2})\vec{a} = \frac{\omega_p^2(z)}{c^2(1-\frac{\sigma\omega_c}{\omega_0})}\left(1 + \frac{r^2}{r_{ch}^2(z)} - \frac{|a|^2}{2\left(1-\frac{\sigma\omega_c}{\omega_0}\right)^2}\right)\vec{a} \tag{6}$$

In this framework, the solution to Eq. (5) is expressed as:

$$\vec{a} = \frac{1}{2}a_0(\hat{x} + i\sigma\hat{y})\exp\left(i\psi - \frac{r^2}{2r_0^2}\right) + \text{c. c.} \tag{7}$$

where the dispersion relation for the propagating laser pulse is given by:

$$k_0(z) = \frac{1}{c}\sqrt{\omega_0^2 - \frac{4c^2}{r_0^2} - \frac{\omega_p^2(z)}{\left(1-\frac{\sigma\omega_c}{\omega_0}\right)}} \tag{8}$$

As the laser pulse propagates through a plasma channel, it perturbs the plasma density and induces a current density. These interactions modify the wave equation, which can be expressed as:

$$(\nabla^2 - \frac{1}{c^2}\frac{\partial^2}{\partial t^2})\vec{a} = \frac{\omega_p^2(z)}{c^2(1-\frac{\sigma\omega_c}{\omega_0})}\left(1 + \frac{r^2}{r_{ch}^2(z)} + \frac{\delta n}{n_0} - \frac{|a|^2}{2\left(1-\frac{\sigma\omega_c}{\omega_0}\right)^2}\right)\vec{a} \tag{9}$$

here $\delta n$ represents the electron density perturbation, which is driven by the ponderomotive force. The relation between $\delta n$ and the laser field is described by:

$$\left(\frac{\partial^2}{\partial t^2} + \frac{4\pi n(r,z)e^2}{m}\right)\frac{\delta n}{n_0} = \frac{c^2}{2}\nabla^2\left(\frac{\vec{a}\cdot\vec{a}}{\left(1-\frac{\sigma\omega_c}{\omega_0}\right)^2}\right) \tag{10}$$

By defining new variables as $\tau = t - \int_0^z \frac{dz'}{v_g(z')}$, $v_g = \frac{c^2 k_0(z)}{\omega_0}$, $\Delta k^2 = -k_0^2 + \frac{\omega_0^2}{c^2} - \frac{\omega_p^2(z)}{\left(1-\frac{\sigma\omega_c}{\omega_0}\right)} - \frac{4}{r_0^2}$, and incorporating Eq. (10), the final equation governing the laser field $a(r,z,\tau)$ is derived as:

$$[\nabla_\perp^2 + \Delta k^2 + \frac{4}{r_0^2} - \frac{\omega_p^2(z)}{c^2\left(1-\frac{\sigma\omega_c}{\omega_0}\right)}\frac{r^2}{r_{ch}^2(z)} + i\frac{\partial k_0}{\partial z}\left(1 - \frac{i}{k_0 v_g}\frac{\partial}{\partial \tau}\right) + 2ik_0\left(1 + \frac{i}{k_0 v_g}\frac{\partial}{\partial \tau}\right)\frac{\partial}{\partial z} + \left(\frac{1}{v_g^2} - \frac{1}{c^2}\right)\frac{\partial^2}{\partial \tau^2} -$$
$$\frac{\omega_p^2(z)}{c^2\left(1-\frac{\sigma\omega_c}{\omega_0}\right)}\left(\frac{\delta n}{n_0} - \frac{|a|^2}{2\left(1-\frac{\sigma\omega_c}{\omega_0}\right)^2}\right)]a(r,z,\tau) = 0 \tag{11}$$

This equation comprehensively describes the evolution of the laser field as it propagates through a plasma channel. These include the group velocity dispersion, which governs the temporal evolution of the laser pulse; plasma density perturbations, which arise due to interactions between the laser and the plasma medium, transverse spatial variations, which capture changes in the laser field across the channel's cross-section, and nonlinear feedback mechanisms. These factors provide a detailed framework for analyzing the complex interplay of linear and nonlinear phenomena in the plasma channel.

## III. Forward Raman scattering in the tapered plasma channel

The reduced equations governing forward Raman scattering and self-modulation are derived analytically, in this section. The derivation is based on the two assumptions, the radius of the plasma channel is significantly larger than $c/\omega_p$, where $\omega_p$ denotes the plasma frequency. And the laser frequency is much greater than $\omega_p$. With considering these assumptions, Eq. (11) simplifies to:

$$[\nabla_\perp^2 + 2ik_0\left(1 + \frac{i}{k_0 v_g}\frac{\partial}{\partial \tau}\right)\frac{\partial}{\partial z} - \frac{\omega_p^2}{c^2\left(1-\frac{\sigma\omega_c}{\omega_0}\right)}\phi]a(r,z,\tau) = 0 \qquad (12)$$

where $\phi = \frac{\delta n}{n_0} - \frac{|a|^2}{2\left(1-\frac{\sigma\omega_c}{\omega_0}\right)^2}$ is defined as a dimensionless term. Reorganizing Eq. (12) yields the following expression:

$$(1 - \frac{i}{k_0 v_g}\frac{\partial}{\partial \tau})[\nabla_\perp^2 - \frac{\omega_p^2}{c^2\left(1-\frac{\sigma\omega_c}{\omega_0}\right)}\phi]a = -2ik_0\frac{\partial a}{\partial z} \qquad (13)$$

In the linear regime, where the growth rate of the instability is much smaller than the plasma frequency, the quasi-static approximation can be applied. Under this, the transverse gradient vector is expressed as $\vec{\nabla}_\perp - \hat{z}\frac{1}{v_g}\frac{\partial}{\partial \tau}$. With this substitution, the equation for electron density perturbation, Eq. (10), is reformulated as:

$$\left(\frac{\partial^2}{\partial t^2} + \omega_p^2\right)\phi = \frac{1}{2}[c^2\nabla_\perp^2 - \omega_p^2](\frac{\vec{a}\cdot\vec{a}}{\left(1-\frac{\sigma\omega_c}{\omega_0}\right)^2}) \qquad (14)$$

These equations provide a detailed theoretical framework for describing the dynamics of forward Raman scattering and self-modulation under the specified plasma conditions. This analytical treatment highlights the interplay between the laser pulse and the plasma, capturing key physical mechanisms essential for understanding these phenomena. To evaluate the growth rate of forward Raman scattering, the vector potential $\vec{a}$ and the electron density perturbation $\phi$ are expressed as follows:

$$a = a_0 + a_+(z,\tau)e^{i\theta} + a_-(z,\tau)e^{-i\theta} \qquad (15)$$

$$\phi = \frac{-|a_0|^2}{2} + [\Phi e^{i\theta} + c.c.] \qquad (16)$$

where the phase is defined as $\theta = \vec{k}\cdot\vec{r} - \omega_p(z)\tau$. Substituting Eqs. (15) and (16) into Eqs. (13) and (14) leads to the following coupled equations for the evolution of $a_+$, $a_-$ and $\Phi$:

$$\frac{\partial}{\partial z}\left(a_+ e^{-i\omega_p\tau}\right) = \frac{-ie^{-i\omega_p\tau}}{2k_0}[1 - \frac{i}{k_0 v_g}\left(\frac{\partial}{\partial \tau} - i\omega_p\right)](\frac{\omega_p^2}{c^2\left(1-\frac{\sigma\omega_c}{\omega_0}\right)}a_0\Phi + k_\perp^2 a_+) \qquad (17)$$

$$\frac{\partial}{\partial z}\left(a_-^* e^{-i\omega_p\tau}\right) = \frac{ie^{i\omega_p\tau}}{2k_0}[1 + \frac{i}{k_0 v_g}\left(\frac{\partial}{\partial \tau} - i\omega_p\right)](\frac{\omega_p^2}{c^2\left(1-\frac{\sigma\omega_c}{\omega_0}\right)}a_0\Phi + k_\perp^2 a_-^*) \qquad (18)$$

$$\frac{\partial \Phi}{\partial \tau} = \left(\frac{-ia_0}{8\omega_p}\right) \frac{(c^2 k_\perp^2 + \omega_p^2)}{\left(1 - \frac{\sigma\omega_c}{\omega_0}\right)^2} (a_+ + a_-^*) \tag{19}$$

Assuming $k_\perp = 0$, the following equation is obtained:

$$\left(\frac{\partial^2}{\partial z \partial \tau} - \frac{\Gamma_{FR}^2}{v_g}\right) \Phi = \frac{\partial}{\partial z} \left(\ln\left(\frac{e^{-i\omega_p \tau}}{\omega_p}\right)\right) \frac{\partial \Phi}{\partial z} \tag{20}$$

where $\Gamma_{FR} = \omega_p^2(z) a_0 / 2\sqrt{2}\omega_0 (1 - \frac{\sigma\omega_c}{\omega_0})$ is the growth rate of forward Raman scattering. For unmagnetized plasmas ($\omega_c = 0$), the growth rate reduces to the result in Ref. [11]. Assuming $\Phi \sim \exp\left[\left(i\delta k - i\frac{\delta\omega}{c}\right) z - i\delta\omega\tau\right]$, Eq. (20) simplifies to:

$$\left(\delta k - \frac{1}{c}\delta\omega\right) \delta\omega = \frac{\Gamma_{FR}^2}{v_g} \tag{21}$$

with the solution for $\delta\omega$ given by:

$$\delta\omega = \frac{1}{2}\{c\delta k + i\sqrt{4\Gamma_{FR}^2 - c^2\delta k^2}\} \tag{22}$$

Thus, the growth rate of forward Raman scattering for a plasma with a linear density profile is:

$$\gamma = \frac{1}{2}\left[4\Gamma_{FR}^2 - \omega_{po}^2 \left(\left(1 + \frac{z}{L}\right)^{\frac{1}{2}} - \frac{ck}{\omega_{po}}\right)^2\right]^{\frac{1}{2}} \tag{23}$$

This equation accounts for amplification mechanisms driven by the energy transfer from the laser to the plasma wave and damping effects arising from frequency mismatches and other dissipative processes. By incorporating position-dependent parameters such as the scale length of the plasma density gradient and the plasma wave frequency, Eq. (23) allows one to predict how the FRS growth rate varies along the laser propagation path. It shows how these factors can enhance or suppress the instability depending on the specific plasma conditions.

### IV. Self-modulation instability in the tapered plasma channel

In this section, the growth rate of the self-modulation instability is analytically derived using Eqs. (17)–(20). By neglecting the operator in Eq. (12) and introducing the following new variables as defined in Ref. [11]:

$$a_\pm = \bar{a}_\pm \exp(\pm i\Delta z) \tag{24}$$

$$\Phi = \bar{\Phi} \exp(i\Delta z) \tag{25}$$

the governing equations can be rewritten as:

$$\frac{\partial \bar{a}_+}{\partial z} = -\frac{ik_p^2 a_0}{2k_0\left(1 - \frac{\sigma\omega_c}{\omega_0}\right)^2} \bar{\Phi} \tag{26}$$

$$(\frac{\partial}{\partial z} + 2i\Delta)a_-^* = \frac{ik_p^2 a_0}{2k_0\left(1-\frac{\sigma\omega_c}{\omega_0}\right)^2}\bar{\Phi} \tag{27}$$

where $\Delta = -k_\perp^2/2k_0$. By insertion of $\hat{a}_\pm = \bar{a}_\pm \exp(\mp i\omega_p\tau)$ and $\hat{\Phi} = \bar{\Phi}\exp(-i\omega_p\tau)$ into the above equations, the following differential equation is obtained:

$$c^2\frac{\partial}{\partial z}\left\{\frac{e^{-i\omega_p\tau}}{\omega_p}\frac{\partial}{\partial \tau}\left(e^{i\omega_p\tau}\left(\frac{\partial}{\partial z} + 2i\Delta\right)\right)\right\}\hat{a}_-^* = ig^2\hat{a}_-^* \tag{28}$$

where the coupling coefficient is defined as $g = a_0 k_\perp/4k_0\left(1-\frac{\sigma\omega_c}{\omega_0}\right)^2$. To evaluate the growth rate of the self-modulation instability, the solution is assumed to have the form $\bar{a}_-^* \sim \exp[(i\delta k - i\frac{\delta\omega}{c})z - i\delta\omega\tau]$. Substituting this expression into the above differential equation yields the dispersion relation:

$$\delta\omega(\delta\omega - c\delta k)^2 = g^2\omega_p^3 \tag{29}$$

By treating $\delta\omega$ as a complex variable, the growth rate of the self-modulation instability is determined. The maximum growth rate is expressed as:

$$\gamma = \frac{\sqrt{3}}{2}\frac{\omega_{p0}}{\left(1-\frac{\sigma\omega_c}{\omega_0}\right)^{\frac{4}{3}}}\left(\frac{a_0 k_\perp}{4k_0}\right)^{\frac{2}{3}} \tag{30}$$

Notably, Eq. (42) reduces to the unmagnetized plasma result presented in Ref. [11] when $\omega_c = 0$. This highlights the impact of the magnetic field on the growth rate of self-modulation instability, demonstrating how the cyclotron frequency modifies the instability dynamics. The result provides a robust analytical framework for understanding and optimizing the self-modulation process in magnetized plasma environments.

## V. Discussion

Tapered channels enable the stable propagation of laser pulses over long distances, which is important for plasma-based accelerators [23]. This channel is proposed to resolve the problem of dephasing in laser-wakefield acceleration [30]. Forward Raman scattering is a parametric instability commonly observed in tapered plasma channels. Stokes and Anti-Stokes sidebands, observed in the spectrum in experimental studies, are generated by the interaction of laser pulse and plasma waves. The intensity of these sidebands, which are downshifted and upshifted from the frequency of laser pulse by the plasma frequency, depends on the growth rate of forward Raman scattering. An increase in the growth rate of this instability amplifies the intensity of forward scattered waves. In the present study, it is assumed that the tapered channel is immersed in a dc magnetic field, which is considered parallel to the wavevector of the laser pulse. The presence of the magnetic field modifies the dispersion relation of electromagnetic waves propagating in plasmas, resulting in the modification of the growth rate of instabilities in plasmas. Self-modulation and forward Raman scattering, which affect laser propagation in tapered plasma channels, are investigated in the present study. Optimized laser and plasma parameters, which are required for the reduction of the growth rate of instabilities and stable propagation of laser pulses in the plasma channel, are investigated in this analysis.

The local growth rate of instability, given by Eq. (23), is plotted against $\hat{z}$ for different wave numbers ($\hat{k}$) in Fig. 2. The curves are confined to specific regions due to the spatial variation of plasma density in tapered plasma channels. As $\hat{k}$ increases, the maximum local growth rate of forward Raman scattering rises. This behavior is governed by the dispersion relation between the laser and plasma waves. Higher $\hat{k}$ values, corresponding to shorter wavelengths, improve phase-matching conditions, enhancing energy transfer between the waves. In tapered plasma channels, the spatial variation of plasma density modifies the local plasma frequency, directly affecting the resonance conditions for Raman scattering. Higher $\hat{k}$ modes are more sensitive to these density gradients, resulting in stronger coupling and increased growth rates in regions of favorable resonance. Furthermore, the slower group velocity of higher $\hat{k}$ modes extends their interaction time within localized plasma regions, amplifying the instability. Collectively, higher wavenumbers enhance resonance conditions, coupling strength, and interaction time, leading to a significant increase in the growth rate of forward Raman scattering.

Figure 3 displays the normalized vector potential logarithmic variations as a function of the propagation distance $z/Z_R$ and time $\tau c/\lambda_p$ for different plasma frequency, where $Z_R$ is the Rayleigh length. Furthermore, the simulation of various plasma frequency effects on the variations in the 3D normalized vector potential is investigated and depicted in Fig. 4. According to the figure, it is clear that as the plasma frequency increases, the number of normalized potential vector maxima increases and the intensity of the scattered waves increases. An increase in plasma frequency signifies a higher electron density within the plasma, which fundamentally alters the dynamics of laser-plasma interactions. Specifically, the electron plasma wave frequency rises, enabling stronger coupling between the laser field and plasma waves, and promoting more efficient energy transfer during parametric processes such as forward Raman scattering (FRS). An increase in plasma frequency facilitates the excitation of a broader spectrum of resonant plasma wave modes. These additional modes result in a more complex distribution of energy within the plasma, leading to an increased number of maxima in the normalized vector potential. This behavior arises from the interference patterns generated by the interaction of the incident and scattered waves with the plasma wave. Moreover, the higher electron density associated with an increased plasma frequency intensifies nonlinear processes. The stronger nonlinear coupling amplifies the excitation of plasma waves, redistributing energy more effectively across the wave modes. As a result, the forward Raman scattering process becomes more pronounced, leading to an increase in both the number of maxima in the normalized vector potential and the overall amount of scattering.

Figure 5 demonstrates the variations of the 3D normalized vector potential versus the normalized $\tau/\tau_L$ and the normalized propagation distance ($z/\lambda_p$) for various external magnetic fields. The observed trends in Figure indicate that the increasing external magnetic field enhances the coupling between the laser and plasma waves, leading to greater energy transfer and amplification of the normalized vector potential. Simultaneously, the stronger magnetic field intensifies the Raman scattering process by amplifying the plasma wave amplitudes and improving the resonance conditions. When an external magnetic field introduces anisotropy into the plasma, electron dynamics are fundamentally altered through the Lorentz force. The magnetic field constrains the transverse motion of electrons, leading to more efficient energy transfer to the plasma waves along the propagation direction of the laser field. The magnetic field's presence alters the plasma's effective dielectric properties, enabling stronger interaction between the laser field and the plasma medium. This modification impacts the dispersion relations of plasma waves and facilitates stronger coupling between the laser field and plasma waves. Raman scattering is a nonlinear parametric process in which the laser propagates in the plasma, causing disturbances in the plasma density and generating plasma waves. The coupling between the plasma waves and the laser leads to the generation of scattered Stokes and anti-Stokes waves. In magnetized plasma, the dispersion relation of plasma waves is influenced by both the plasma frequency ($\omega_p$) and the cyclotron frequency ($\omega_c$). The magnetic field increases the oscillatory response of electrons in the direction of the field. This amplification of electron oscillations strengthens the coupling between the laser field and the plasma waves, resulting in higher plasma wave amplitudes and Raman scattering with higher intensity.

The impact of the polarization state on variations in the number of e-folding in forward Raman scattering is depicted in Figure 6. The findings in Figure highlight the superior performance of linearly polarized laser fields in driving forward Raman scattering compared to right-handed and left-handed circularly polarized fields. This behavior can be attributed to the distinct ways in which polarization affects the coupling between the laser field and plasma waves, influencing the efficiency of energy transfer and the growth of plasma wave amplitudes. In linear polarization, the electric field oscillates in a fixed plane, resulting in coherent and directed electron oscillations along the polarization axis. This alignment leads to more efficient energy transfer, and enhances the coupling efficiency between the laser field and the plasma waves, leading to stronger excitation of plasma oscillations and, consequently, a higher number of e-foldings. In addition, the laser intensity is concentrated along a fixed direction, amplifying the plasma wave excitation in that direction. This directional intensity enhances the nonlinear interaction driving forward Raman scattering. In contrast, for circular polarizations, the electric field vector rotates in a circular path, causing a time-varying phase difference between the laser field and the plasma waves. This variation disrupts the resonance conditions required for optimal energy transfer, reducing the parametric growth rate and the corresponding number of e-foldings. Moreover, the helical motion distributes the electron response across a wider spatial domain and distributes the laser intensity more uniformly across all directions. This uniformity reduces the effective intensity available to drive the forward Raman scattering process, thereby resulting in fewer e-foldings. As a result, the number of e-foldings in FRS is lower for circular polarization.

Figure 7 displays the role of normalized vector potential on variations of the number of e-folding in forward Raman scattering versus the scaled wave number $\hat{k}$. Figure underscores the pivotal role of the normalized vector potential $a_0$ in determining the efficiency and dynamics of forward Raman scattering. The observed decrease in the number of the e-folds with reducing $a_0$ is attributed to a combination of diminished energy transfer efficiency, lower parametric growth rates, and restricted resonance conditions. The normalized vector potential $a_0$ is a dimensionless parameter that quantifies the strength of the laser field relative to the plasma's natural electromagnetic field. A higher value of $a_0$ corresponds to a stronger laser field, which induces more robust oscillatory motion in plasma electrons, enhancing the efficiency of energy transfer from the laser field to the plasma waves, and drives larger plasma wave amplitudes, which directly increase the parametric gain and the number of e-foldings. In contrast, a reduction in $a_0$ weakens the coupling between the laser field and plasma waves, resulting in lower plasma wave amplitudes and smaller e-fold numbers. Forward Raman scattering arises from a parametric instability, where the growth rate is strongly dependent on the laser field intensity, which is proportional to $a_0^2$. A higher normalized vector potential leads to an enhanced growth rate of the plasma waves, promoting exponential amplification over a broader range of the scaled wave number $\hat{k}$. When $a_0$ decreases, the range of $\hat{k}$ values that satisfy the resonance conditions narrows, further limiting the parametric amplification, the growth rate diminishes, weakening the instability and reducing the extent of exponential amplification, which manifests as a decrease in the number of the e-folds.

The effect of plasma frequency on variations of the number of e-folding in forward Raman scattering versus the scaled wave number $\hat{k}$ is investigated and depicted in Fig. 8. As the figure shows, the amount of folding increases as the plasma frequency increases. This trend is explained by the enhanced resonance conditions, higher parametric growth rates, improved phase matching, and increased plasma susceptibility associated with higher plasma frequency. As $\omega_p$ increases, the plasma becomes more responsive to the electromagnetic field of the laser, enhancing the resonance between the laser field and the plasma waves. This means that at higher plasma frequencies, the instability develops more rapidly, leading to faster and stronger amplification of the plasma waves. And, this enhanced growth directly contributes to the increase in the number of e-foldings, as the plasma waves experience greater exponential growth within the same propagation distance. In addition, higher plasma frequencies modify the dispersion relation in a manner that facilitates a broader range of scaled wave numbers $\hat{k}$ for which efficient wave coupling occurs. Improved phase matching between the laser field and plasma waves, enabling continuous energy transfer and wave amplification. This broader coupling range and improved phase matching contribute to the observed increase in the number of e-foldings with higher.

In Figure 9, the role of the cyclotron frequency on variations of the number of e-folding in forward Raman scattering versus the scaled wave number $\hat{k}$ is showcased across other distinct sets of parameters. As the figure shows, the number of folds increases with increasing external magnetic field or cyclotron frequency. Figure reveals that the number of e-foldings in forward Raman scattering increases with the cyclotron frequency, or equivalently, with a stronger external magnetic field. This behavior is attributed to enhanced electron confinement, broader resonance conditions, improved phase matching, higher parametric growth rates, and strengthened nonlinear coupling effects introduced by the magnetic field. The cyclotron frequency ($\omega_c$) represents the angular frequency at which charged particles, such as electrons, gyrate around magnetic field lines. The magnetic field confines electron motion more effectively, increasing their oscillatory interaction with the laser field. This confinement leads to more efficient coupling of the laser energy into plasma waves, thereby amplifying the growth of forward Raman scattering and increasing the number of e-foldings. In addition, the magnetic field increases the anisotropy of the plasma, facilitates better phase matching between the laser field and plasma waves, strengthens the nonlinear coupling between the laser field, and expands the range of scaled wave numbers ($\hat{k}$) over which efficient wave coupling occurs, resulting in stronger amplification of plasma waves. This anisotropy supports more efficient wave excitation and amplification, resulting in a higher number of e-foldings.

Figure 10 illustrates the impact of the plasma and cyclotron frequencies on the variations of the growth rate of self-modulation versus $a_0$. The figure demonstrates that the growth rate of self-modulation increases with rising plasma and cyclotron frequencies. This enhancement is driven by stronger resonance conditions, improved energy coupling, and intensified nonlinear interactions enabled by plasma and cyclotron frequencies. The plasma frequency ($\omega_p$) governs the collective oscillatory behavior of electrons in the plasma. The enhanced electron oscillations lead to a more robust resonance between the laser field and the plasma, facilitating the growth of self-modulation. Higher $\omega_p$ values result in larger plasma wave amplitudes, which amplify the growth rate of self-modulation. The cyclotron frequency ($\omega_c$) represents the angular frequency of electron gyration in the presence of an external magnetic field. The magnetic field anisotropically modifies the plasma dispersion relations, enabling stronger coupling between the laser and plasma waves. Higher $\omega_c$ values strengthen nonlinear interactions, such as parametric coupling and wave-wave interactions, which contribute to a higher self-modulation growth rate. The simultaneous increase in both plasma and cyclotron frequencies creates synergistic effects that further enhance the self-modulation growth rate. Together, these parameters optimize the coupling between the laser field and plasma waves, resulting in more efficient energy exchange and a higher growth rate. Additionally, at sufficiently high values of $a_0$, the growth rate may approach a saturation point due to nonlinear damping or wave-breaking, but the overall enhancement by plasma and cyclotron frequencies remains evident.

## VI. Conclusions

This study analytically and numerically investigates forward Raman scattering and self-modulation instability, which are prominent phenomena in laser-plasma interactions. The governing equations for the propagation of circularly polarized lasers in a tapered magnetized plasma are derived. Partial differential equations describing forward Raman scattering and self-modulation instability are formulated and solved through both analytical and numerical methods. The influence of laser intensity, polarization, cyclotron frequency, and plasma density on the growth rates of these instabilities is systematically examined. Numerical results demonstrate that higher laser intensities and plasma densities significantly enhance the growth rates of forward Raman scattering and self-modulation instability. Furthermore, increased magnetic field strength is observed to augment the instability growth rate for right-handed circularly polarized (RHCP) lasers while reducing it for left-handed circularly polarized (LHCP) lasers. Enhanced growth rates of forward Raman scattering amplify the Stokes and anti-Stokes sidebands in the spectrum, indicating that a larger fraction of the incident laser energy is scattered by the plasma medium. Similarly, the amplification of self-modulation instability accelerates the transformation of the laser pulse envelope into a modulated structure. Optimal selection of laser and plasma parameters can mitigate the growth rates of these instabilities, which is critical for the efficient design of plasma-based accelerators. These findings provide valuable insights for controlling and optimizing laser-plasma interactions in advanced accelerator technologies.

## Acknowledgment

This research did not receive any specific grant from funding agencies in the public, commercial, or not-for-profit sectors.

**List of Figures & Captions**

**Fig. 1.** A schematic depiction of a laser pulse focused within a tapered plasma channel.

**Fig. 2.** Growth rate of forward Raman scattering as a function of $\hat{z}$, with the parameters: $a_0 = 0.01$, $\lambda_0 = 532nm$, $\omega_p = 0.1\omega_0$, $\omega_c = 0.1\omega_0$ for (a) $\hat{k} = 1.61$, (b) $\hat{k} = 1.71$, (c) $\hat{k} = 1.81$, and (d) $\hat{k} = 1.91$.

**Fig. 3.** Variations of the normalized vector potential logarithmic variations as a function of the propagation distance $z/Z_R$ and time $\tau c/\lambda_p$ for various plasma frequency.

**Fig. 4.** Variations of the 3D normalized vector potential versus the normalized $\tau/\tau_L$ and the normalized propagation distance $(z/\lambda_p)$ for various plasma frequencies, $\lambda_0 = 532nm$, $\hat{k} = 1.61$, $\omega_c = 0.1\omega_0$, and $\sigma = 1$.

**Fig. 5.** Impact of external magnetic field or cyclotron frequency on variations of the 3D normalized vector potential versus the normalized $\tau/\tau_L$ and the normalized propagation distance $(z/\lambda_p)$ for $\lambda_0 = 532nm$, $\hat{k} = 1.61$, $\omega_p = 0.1\omega_0$, and $\sigma = 1$.

**Fig. 6.** Variations of the of the number of e-folding in forward Raman scattering versus the scaled wave number $\hat{k}$, for various polarizations states, $\lambda_0 = 532nm$, $\omega_p = 0.1\omega_0$, $\omega_c = 0.1\omega_0$, and $a_0 = 0.01$.

**Fig. 7.** The role of normalized vector potential on variations of the number of e-folding in forward Raman scattering versus the scaled wave number $\hat{k}$, $\lambda_0 = 532nm$, $\omega_p = 0.1\omega_0$, $\omega_c = 0.1\omega_0$, and $\sigma = 1$.

**Fig. 8.** The effect of plasma frequency on variations of the number of e-folding in forward Raman scattering versus the scaled wave number $\hat{k}$, $\lambda_0 = 532nm$, $\omega_c = 0.1\omega_0$, $a_0 = 0.01$ and $\sigma = 1$.

**Fig. 9.** Variations of the number of e-folding in forward Raman scattering versus the scaled wave number $\hat{k}$, for different cyclotron frequencies, $\lambda_0 = 532nm$, $\omega_p = 0.1\omega_0$, $a_0 = 0.01$ and $\sigma = 1$.

**Fig. 10.** Variations of the growth rate of self-modulation versus $a_0$ for various plasma and cyclotron frequencies where $\lambda_0 = 532nm$, $k_\perp = \frac{0.2\,\omega_p}{c}$ and $\sigma = 1$.

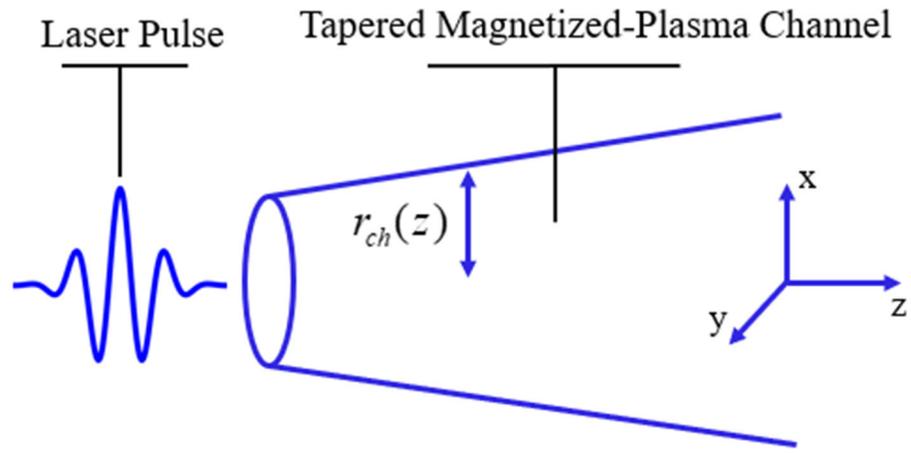

**Fig. 1.**

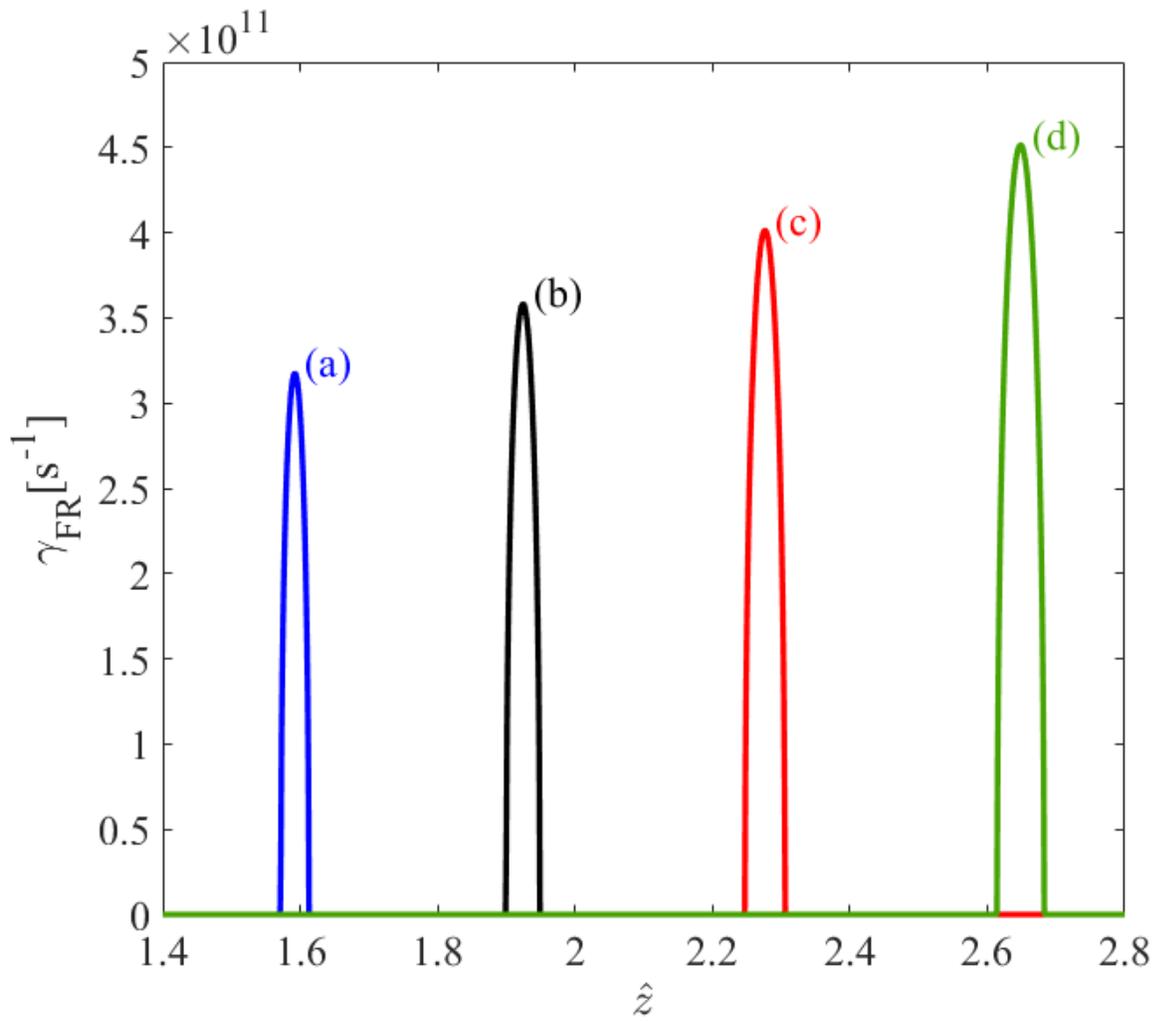

**Fig. 2.**

(a)

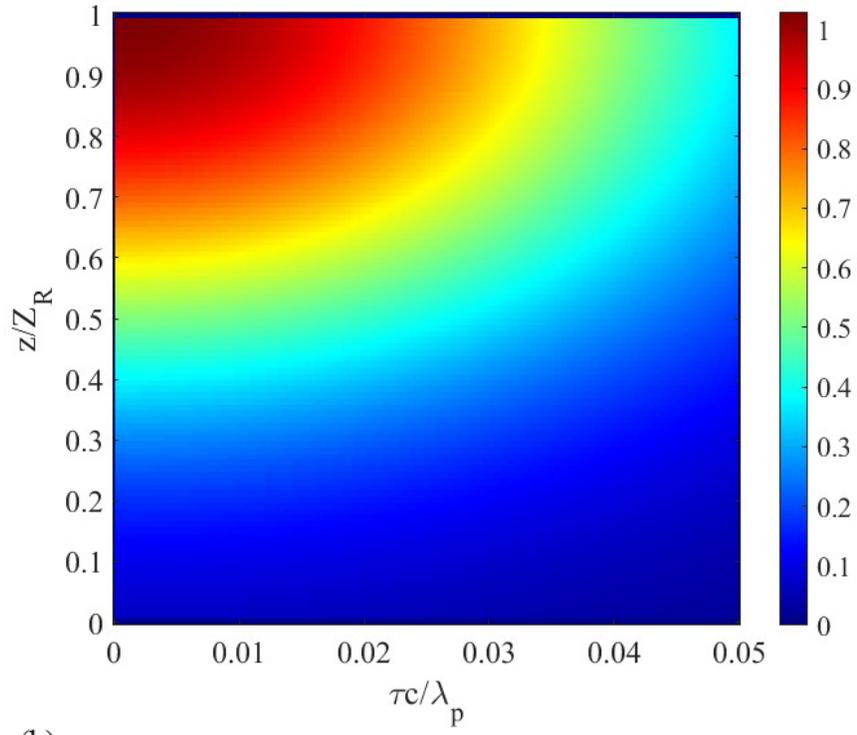

(b)

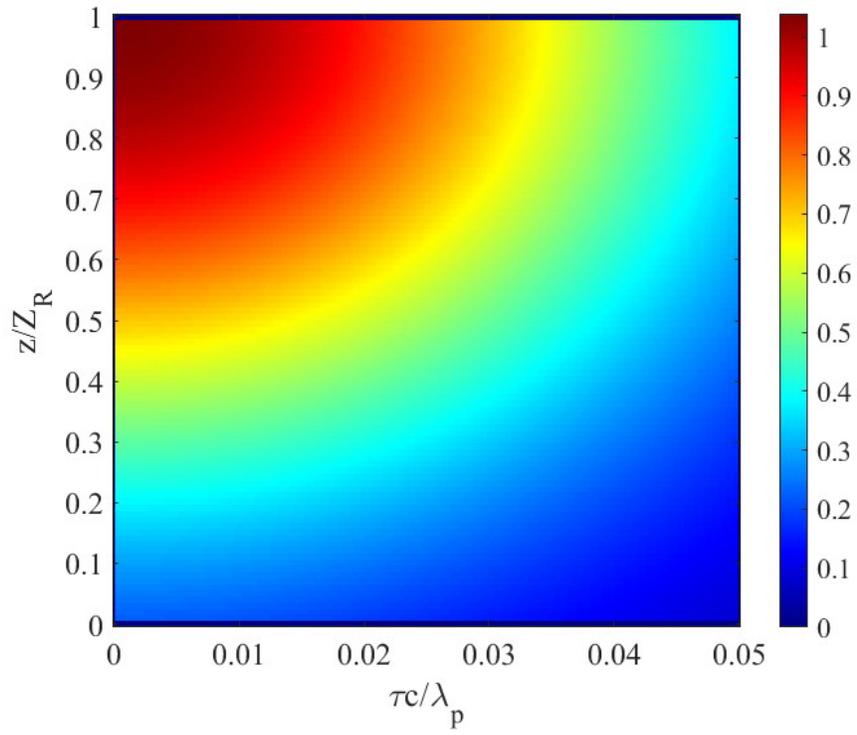

**Fig. 3.**

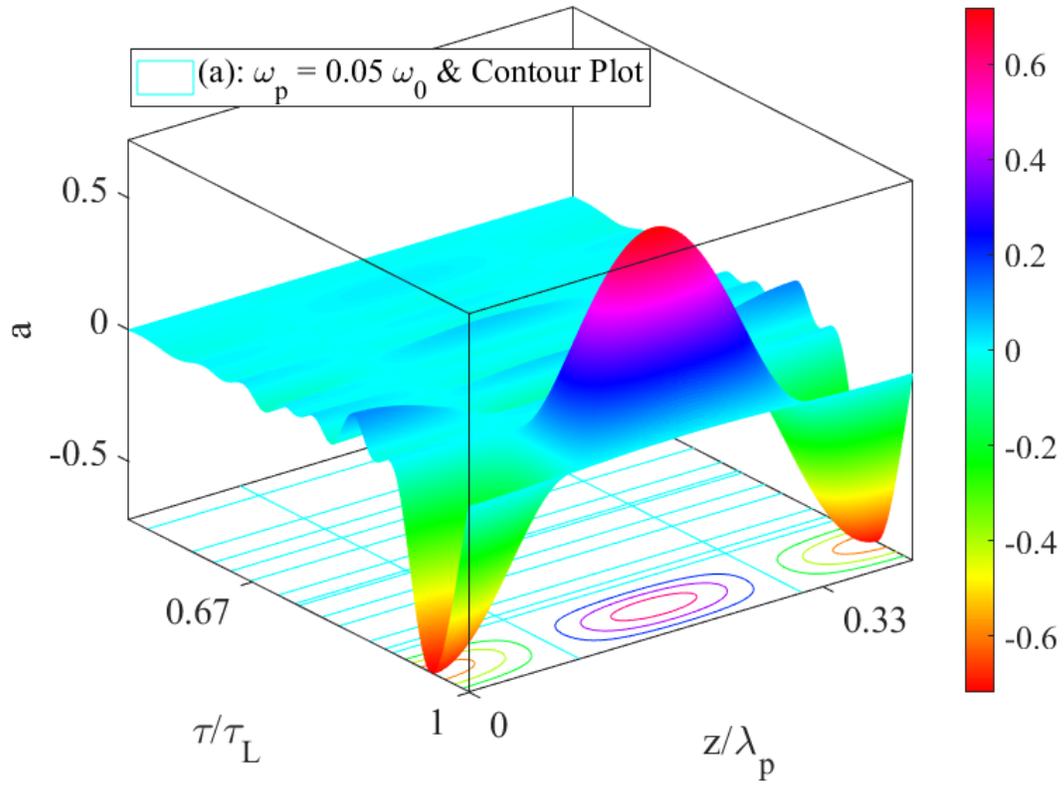
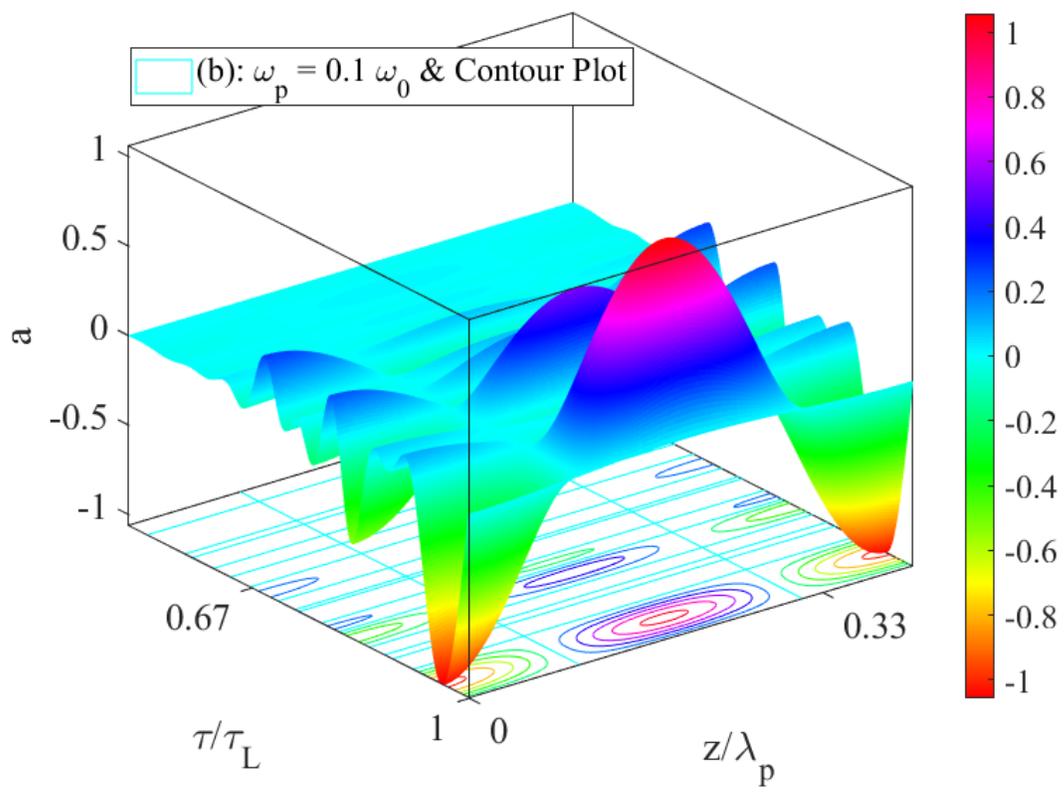

**Fig. 4.**

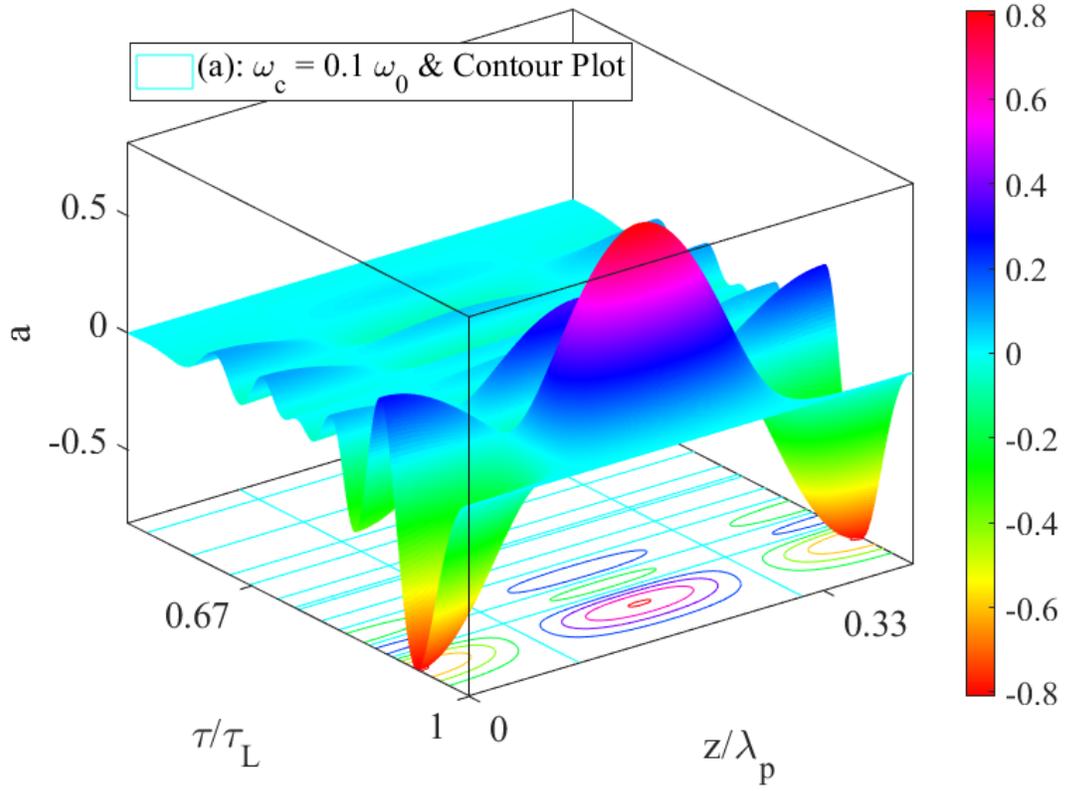
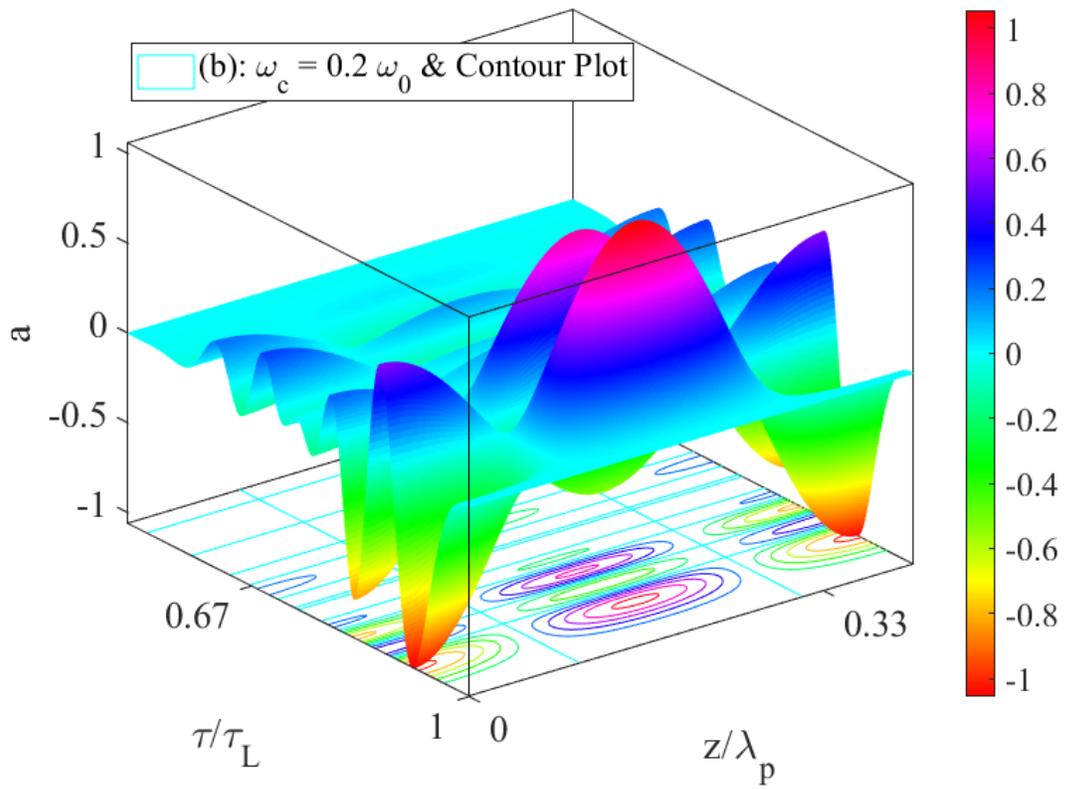

**Fig. 5.**

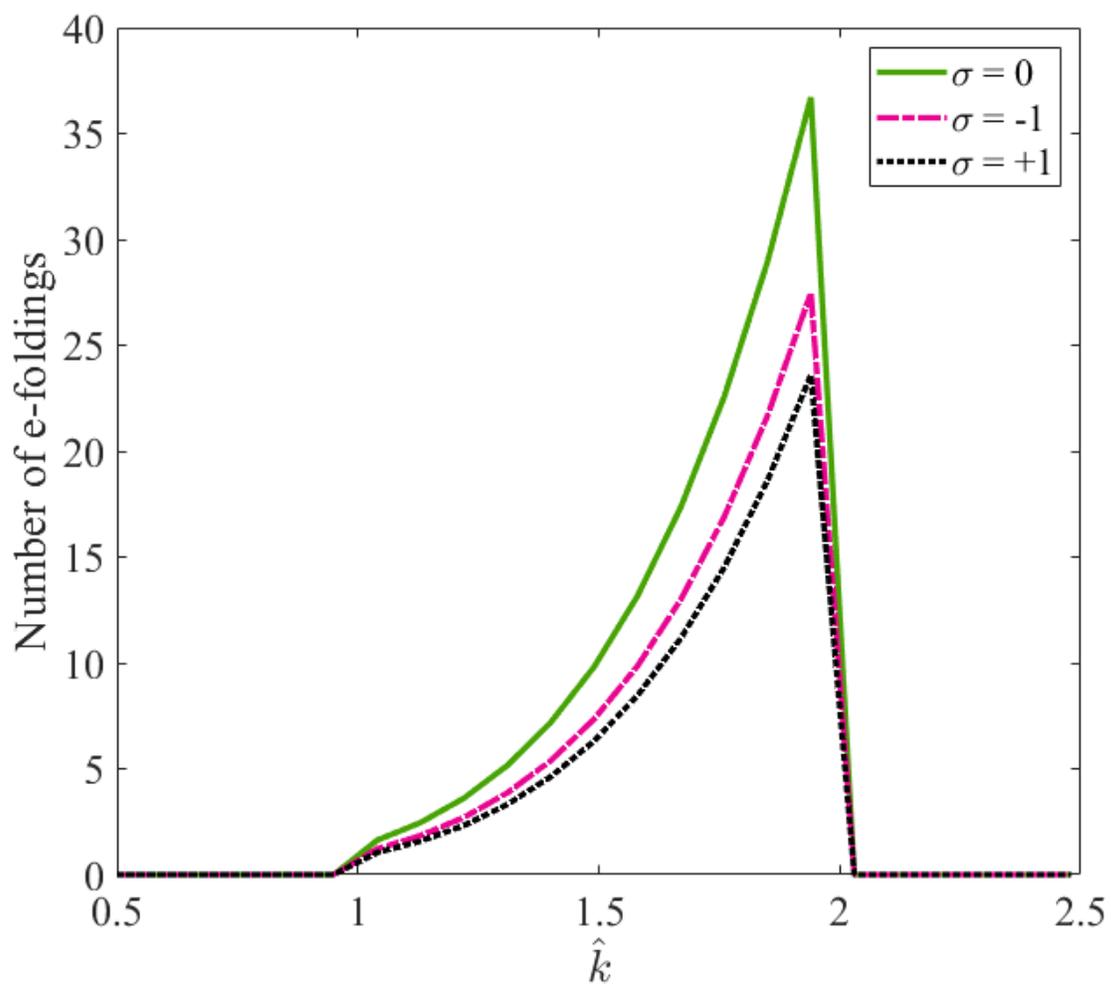

**Fig. 6.**

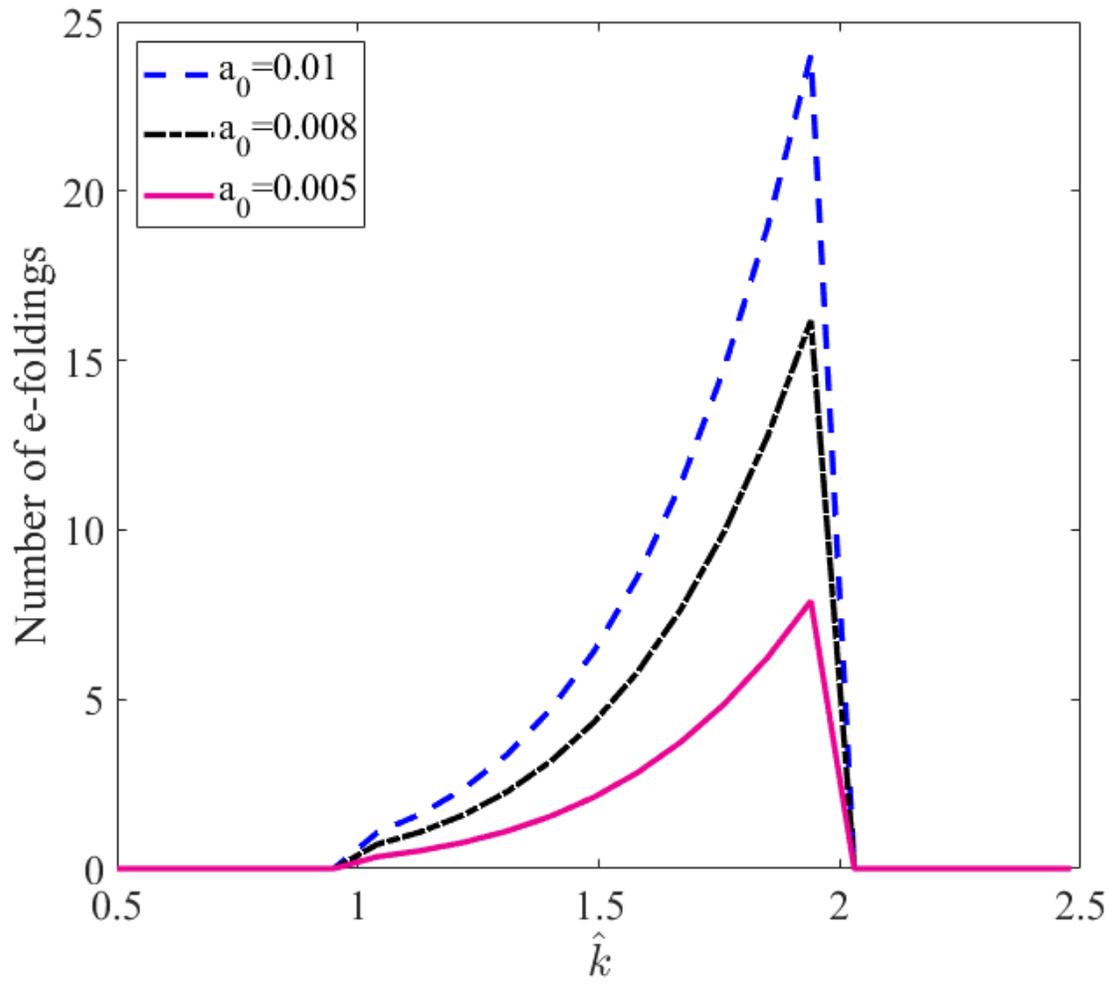

**Fig. 7.**

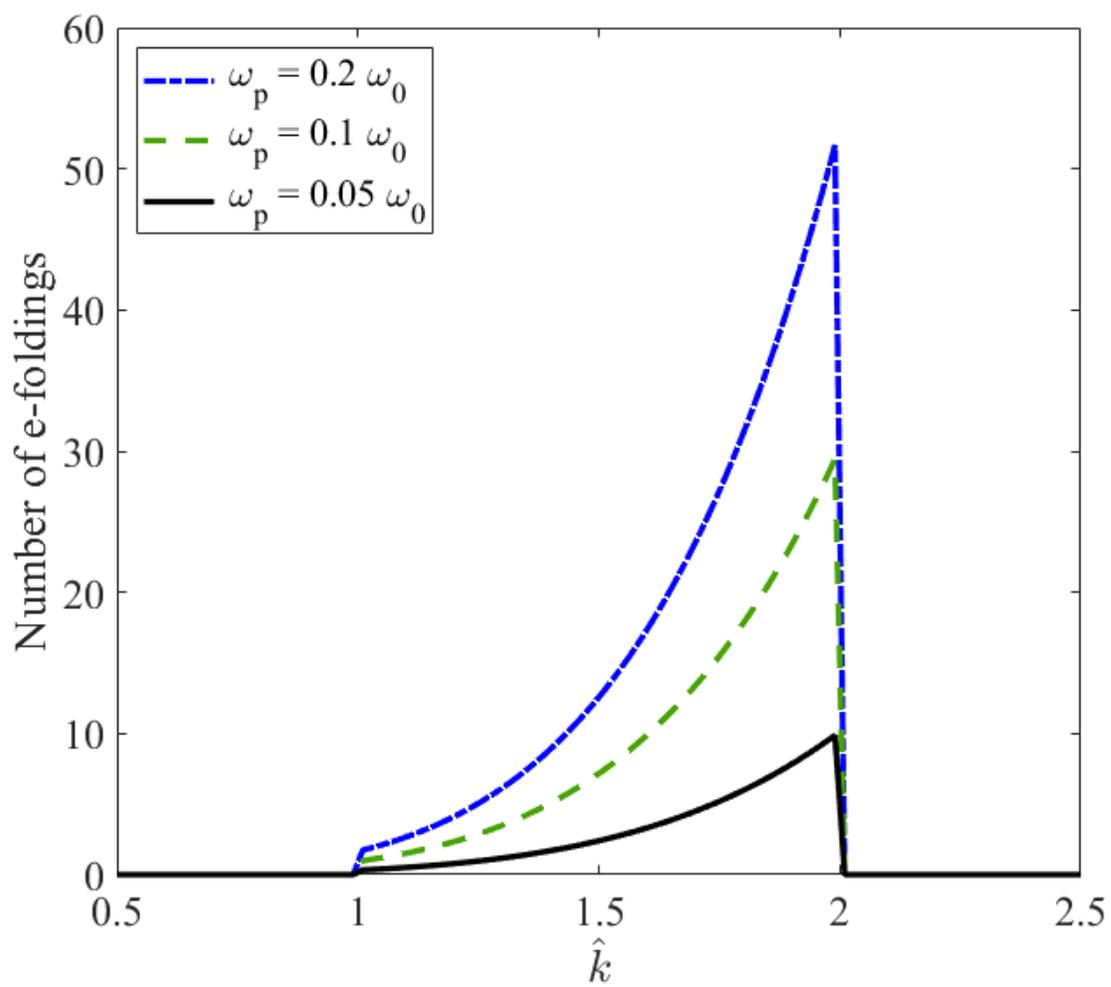

**Fig. 8.**

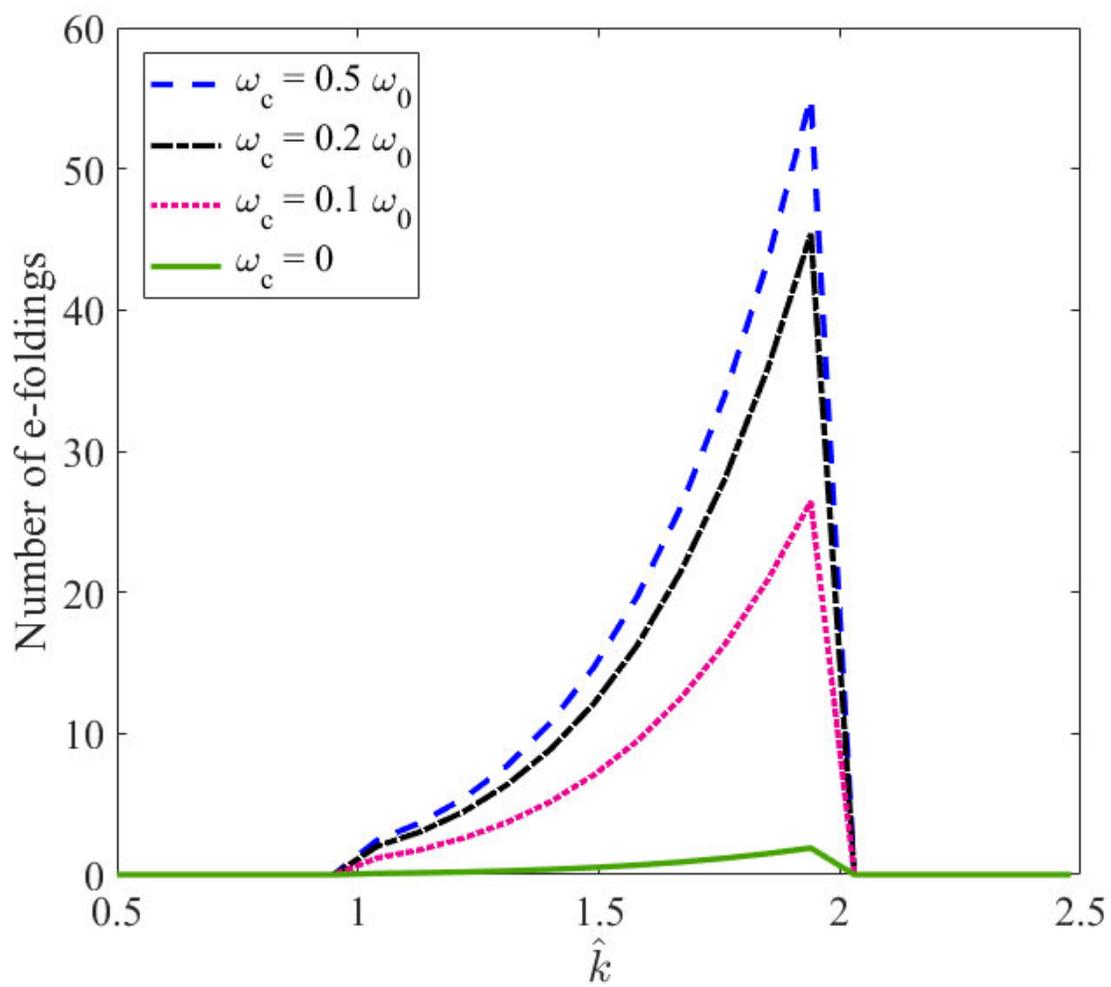

**Fig. 9.**

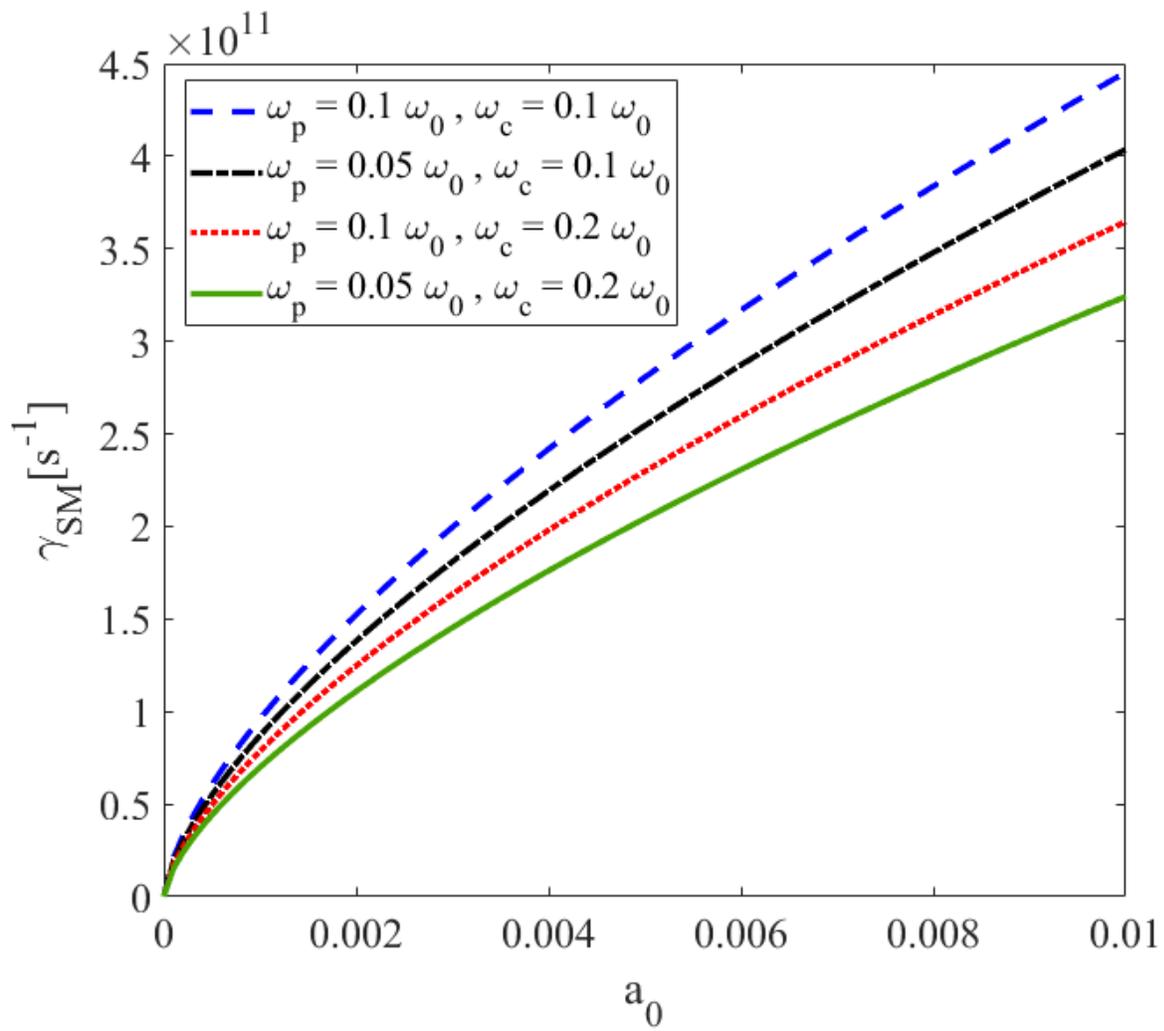

**Fig. 10.**